\documentclass[10pt,journal]{IEEEtran}
\usepackage{float}
\usepackage{multirow}
\usepackage{amsmath,amssymb,amsfonts}
\usepackage{algorithmic}
\usepackage{graphicx}
\usepackage{array}
\usepackage{tabularx}
\usepackage{textcomp}
\usepackage{xcolor}
\usepackage{gensymb}
\usepackage{mathtools}
\usepackage{soul}
\usepackage{booktabs}
\usepackage{comment}
\usepackage{mdwtab}
\DeclarePairedDelimiter\ceil{\lceil}{\rceil}
\DeclarePairedDelimiter\floor{\lfloor}{\rfloor}
\newcolumntype{Y}{>{\centering\arraybackslash}X}
    
\usepackage[style=numeric-comp,subentry,firstinits=true, maxbibnames=3,maxcitenames=3,backend=biber,sorting=none, doi=false,isbn=false, url=false,eprint=false]{biblatex}
\begin{document}
\title{Channel Tiling for Improved Performance and Accuracy of Optical Neural Network Accelerators}

\author{Shurui~Li,
        Mario~Miscuglio,
        Volker~J.~Sorger,
        and~Puneet~Gupta
\IEEEcompsocitemizethanks{\IEEEcompsocthanksitem Shurui Li and Puneet Gupta are with the Department
of Electrical and Computer Engineering, University of California, Los Angeles, CA, USA, 90095.\protect\\
E-mail: shuruili@ucla.edu; puneetg@ucla.edu
\IEEEcompsocthanksitem Mario Miscuglio and Volker J. Sorger are with the Department of Electrical and Computer Engineering, George Washington University, DC, USA, 20052.\protect\\
E-mail: mmiscuglio@email.gwu.edu, sorger@email.gwu.edu}
}

%
\IEEEtitleabstractindextext{%
\begin{abstract}
Low latency, high throughput inference on Convolution Neural Networks (CNNs) remains a challenge, especially for applications requiring large input or large kernel sizes. 4F optics provides a solution to accelerate CNNs by converting convolutions into Fourier-domain point-wise multiplications that are computationally 'free' in the optical domain. 
However, existing 4F CNN systems suffer from the all-positive sensor readout issue, making the implementation of a multi-channel, multi-layer CNN not scalable or even impractical. In this paper, we propose a simple channel tiling scheme for 4F CNN systems that utilizes the high resolution of 4F systems to perform channel summation inherently in the optical domain before sensor detection, so the outputs of different channels can be correctly accumulated. Compared to state of the art, channel tiling gives similar accuracy, significantly better robustness to sensing quantization error (33 percent improvement in required sensing precision) and noise (10dB reduction in tolerable sensing noise), 0.5X total filters required, 10-50X+ throughput improvement and 3X+ reduction in required output camera resolution/bandwidth. Not requiring any additional optical hardware, the proposed channel tiling approach addresses an important throughput and precision bottleneck of high-speed, massively-parallel optical 4F computing systems.
\end{abstract}
}

\maketitle

\IEEEdisplaynontitleabstractindextext

%
\IEEEpeerreviewmaketitle

\section{Introduction}
Convolutional neural networks (CNNs) have proliferated in image classification and computer vision. Over the years, CNNs have become increasingly larger and deeper, making it harder to deploy such networks on traditional electronic machines due to convolution's high computation complexity. Although many efforts have been made on both algorithms and hardware to speed up the convolution process, inference of large CNNs, especially those with high-resolution inputs, is still computationally prohibitive. There is renewed and growing interest in optical/photonic computation hardware for CNN inference acceleration, due to their low compute latency (essentially time of flight of light) and the potential to support large parallelism \cite{photonicnnsurvvey}.

Current photonic CNN accelerators can be roughly classified into two main categories: (1) on-chip implementations using photonic devices including Mach-Zehnder Interferometers \cite{mitopticcnn} and micro-ring resonators \cite{pcnna,mrr2019,photonicmac_2019photonic}; and (2) free-space 4F system, using spatial light modulators (SLM) and phase masks \cite{slm4f,standford4f,opticalrelu,miscuglio2020optica}. 

On-chip photonic implementations usually have a high clock speed (in the GHz range), but the amount of parallelism is far less than the 4F system. The performance of on-chip photonic implementations is usually bounded by digital-to-analog converters' relatively low operation frequency, and they might not be efficient when dealing with high-resolution inputs. Scaling is another issue of on-chip photonic implementations since the photonic hardware scale significantly slower than semiconductors.

In contrast, free-space 4F systems offer massive parallelism due to the high resolution of SLMs and phase masks, as well as efficient convolution computation using the well-known 4F theory, which states that convolution in the space domain is equivalent to point-wise multiplication in the Fourier domain. The 4F system can be implemented using two Fourier lenses, an input source, a device for multiplication and an output sensor. Fourier transform, point-wise multiplication and inverse Fourier transform in the 4F system is essentially constant time (the time of flight of light). Earliest optical correlators based on the 4F theory date back to the 1960s. Recently advancements in CNNs have renewed interest in 4F systems to speed-up the 'expensive' convolution process. Fast SLMs and fast cameras still remain a challenge despite substantial improvements for optical 4F computing. 

In this paper, we focus on improving the accuracy and performance for high-speed, high-resolution optical 4F computing systems for neural network acceleration by proposing a simple tiling method to accumulate the convolution results of all channels of a filter inherently in the optical domain, before the non-linearity applied by photodetectors/cameras. 
The main contributions of this work are summarized as follows.
\begin{itemize}
    \item We provide scalability analysis and performance estimation of free-space 4F systems for CNN acceleration, which indicate that 4F systems have potential to outperform GPUs with advanced hardware and proper algorithm and system configuration.
    \item We propose a channel tiling approach which allows filters to have negative weights, an implicit non-linear activation by the camera/photodetector and optical domain accumulation of channels. We further combine channel tiling with input or filter tiling to fully leverage available optical parallelism and achieves \emph{10-50X} utilization improvements compares with other approaches.
    \item Our results indicate that channel tiling dramatically improves network accuracy (\emph{36 percent} points on CIFAR10 dataset running VGG16 network) compared to alternative tiling approaches. 
    \item We show that channel tiling can be done without any changes to optical hardware itself unlike optical alternatives to preserve sign and it can have (at least) \emph{2X} throughput advantage over the recently proposed pseudo-negative approach.
    \item We show that channel tiling can reduce the photodetector/camera resolution requirements by orders of magnitude alleviating the main bottleneck of 4F computing systems.
    \item We further show that channel tiling is much more robust to camera's quantization error and photo-detection noise (1 percent vs. \emph{15 percent} accuracy drop compared to state of the art with 8-bit camera and 20dB SNR). This allows for channel tiling to support much faster and lower bit-depth cameras.  
\end{itemize}

\section{A Primer on 4F Optical Computing Systems}
\begin{figure}[h]
    \centering
    \includegraphics[width=\linewidth]{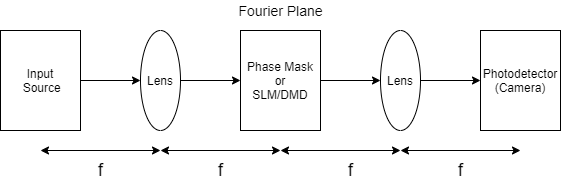}
    \caption{Illustration of optical 4F system.}
    \label{fig:4fsystem}
\end{figure}
\subsection{An Overview of Optical 4F Computing System}
4F optical system usually consists of 4 parts: input source, two Fourier lenses, filter light modulator and output sensor. Figure \ref{fig:4fsystem} shows the high-level system diagram of the 4F optical computing engine. The input source usually includes a laser emitter to generate coherent light and an SLM to encode the input by modifying the light intensity. Then the encoded light passes through the first Fourier lens to perform Fourier transform. The Fourier transformed light signal is projected onto the filter light modulator (Fourier plane), where it actively or passively modulates the light to conduct point-wise multiplication in the Fourier domain. Afterward, the light signal that contains multiplication results passes through the second Fourier lens for inverse Fourier transform. Finally, it is captured by a high-speed camera and enters into the electronic domain.

Most published works \cite{standford4f,slm4f} use phase masks as the filter light modulator to modulate the light in the Fourier plane passively. SLM and DMD (digital micromirror device, a special type of SLM) can also be used as filter light modulators to actively modulate the light, hence providing programmability. \cite{miscuglio2020optica} recently demonstrates the first optical 4F CNN implementation with real-time programmable filters, which uses high-speed DMD for both input and filter generation. For passive approaches, multiplication can be implemented with zero latency and power, but with no flexibility since the weights are fixed. In contrast, active approaches have more flexibility (weights can be modified) and can hence easily scale to large multi-layer networks, albeit at the cost of latency and power overheads.

\subsection{Introduction of 4F Computing System Hardware}\label{hardwareintro}
The field of optical neural networks including 4F based systems is gaining interests and becoming increasingly active. However, it is still at its beginning stage with huge potentials as well as many constraints and issues. Thus many works at this stage focus more on proof of concept rather than build a full, working system that outperforms state-of-art electronic systems. Even though these proof of concept works do not demonstrate better performance than electronics, optical computing systems still shows promise to overtake electronic counterparts, especially with the help of ongoing development and evolution of advanced optic devices and materials. In this subsection we will provide brief introduction to the hardware that could be used in optical 4F systems and their performance estimation.

\subsubsection{Spatial Light Modulator}\label{SLMintro}
Spatial light modulator (SLM) is a core component of the 4F system. It modulates the amplitude and/or phase of incoming light according to the programmed values hence can be used for both input and filter generation. There are several types of light modulators that are widely available: liquid crystal spatial light modulator and micromirror arrays (MMA) which can be further classified into digital MMA (DMD) and analog MMA.  

{\em Liquid crystal SLM} can modulate the phase and/or the intensity of light directly by adjusting the cell's refractive index. Though they can offer high resolution such as 4K \cite{HOLOEYESLM} and can modulate both amplitude and phase of incoming light directly \cite{slmmodulatingref}, even high-speed SLMs operate below one kilohertz \cite{700kslm,400kslm} (due to the nature of liquid crystal, adjusting the phase takes a long time). The low operating frequency makes liquid crystal SLM not suitable for CNN acceleration.

To overcome liquid crystal SLM's low operating frequency, ongoing researches are trying to use alternative materials to replace liquid crystal, enabling fast operating SLM. For example, \cite{Peng19ghzslm} proposed a phase-only SLM architecture that uses microcavities with barium titanate to achieve GHz operation frequency with high-pixel resolution. If the concept can be materialized, 4F systems could potentially operate in the GHz regime and SLM will no longer be the system's bottleneck.

{\em MMAs} modulate light intensity at high speed by flipping their micromirrors to deflect input light \cite{DMDconcept,dmdtheory}. Compares to SLM, the advantage of MMA is its high operating frequency, though it cannot modulate phase directly \cite{oamDMD,superpixeldmdmodulation}. For {\em DMD}, each pixel contains a mirror and a memory unit, and the mirror flips according to the value stored in memory to let the light either pass or deflect away. Gray value modulation is implemented using a time-multiplexing technique similar to pulse width modulation, where the length of bit (duration where the mirror is on) is weighted by its corresponding power of two. The modulation result is measured by the averaging intensity over the entire multi-bit duration. Current commercially available DMD resolution can scale up to 4K \cite{1080p23kDMD,4KDMD}, with a nominal operating frequency of 20 to 30 KHz for binary mode \cite{1080p23kDMD}.   

Unlike DMDs which have a fixed micromirror tilt angle to pass or deflect the light, the tilt angle of {\em analog MMA's} micromirrors can be adjusted by voltage according to the desired intensity. The main idea behind such design is MMA could be treated as an optical grating, hence by adjusting the mirror tilting angle slightly, the intensity of zeroth diffraction order is also adjusted \cite{IPM1mhzdmd,analogmmaref1}. Therefore analog MMA naturally supports multi-bit mode without sacrificing operating frequency. Analog MMAs require digital to analog converters (DAC) to encode pixel values into voltages applied to MMA.\footnote{8bit+ resolution, MHz speed DACs are well behind the state of the art \cite{11GSDAC,12GSDAC} and are not expected to be the bottlenecks for MMAs.} Analog MMAs are under active research and can achieve much higher switching speeds than commercial DMDs \cite{IPM1mhzdmd,2.3MHzDMD}. For instance, \cite{2.3MHzDMD} (11M pixels, 2.3MHz) and \cite{IPM1mhzdmd} (2.2M pixels, 1MHz) have demonstrated high speed, high resolution MMAs. 

Most spatial light modulator can either modulate the phase or the amplitude of incoming light, while only very few can modulate both simultaneously. However, to accurately compute convolution using 4F system, complex weight representation is desired, hence both amplitude and phase need to be modulated. Since the support of complex representation using either phase-only or amplitude-only SLM is also beneficial to many other use cases, extensive researches has been conducted to solve this problem. \cite{arbitrarycomplexusingphaseslmnature,simultaneouscomplexusingphaseslmnature} propose different methods to enable complex modulation using phase-only SLM while complex modulation using amplitude-only SLM can be addressed using the concept of Mach-Zehnder Interferometer \cite{mziref} or spatial encoding \cite{superpixeldmdmodulation,phaseencodedmd,complexbinarydmd}. Thus given either phase-only or amplitude-only SLMs, full complex modulation can be achieved using the cited techniques, hence enabling accurate convolution computation.

\subsubsection{High-speed Camera}\label{cameraintro}
For high-speed 4F optical systems, the output sensing is usually the performance bottleneck due to the performance gap between high-speed SLM and high-speed camera. In such systems, the inputs are usually generated by high-resolution, high-speed SLM/DMD, while the multiplication is carried out using either another SLM or phase mask. As discussed, commercial SLMs can operate in 4K resolution at 20-30 KHz \cite{4KDMD}, with several research SLMs going to MHz/GHz regime. The operation frequency of such SLM is much higher than the state-of-art commercial 4K high-speed camera which operates at roughly 1 KHz \cite{phantom4kcam}. The main performance bottleneck of high-speed cameras is the readout time, which scales with the camera's resolution. The operating frequency of high-speed cameras can match or exceed SLM if the resolution goes down to around 1K, as there are high-speed framing cameras that can operate in the MHz or even GHz range \cite{SpecializedImaging7M,SpecializedImaging1B}. But it also means the SLM's high resolution cannot be fully utilized since normally the camera and SLM should have the same resolution.

\subsubsection{Overall System}
Optical 4F system has many variations, and the overall system capability and performance depend on the system setup and devices used. In this paper we consider a programmable 4F system which uses SLMs for both input and filter modulation, and full complex modulation capability is assumed. For system performance, based on previous analysis, a 4K SLM operating at 2MHz is assumed for the system setup, which is realizable in the near future. For simplicity of performance analysis and comparison, we assume the camera can operate at the same frequency and resolution as SLMs used in the system thus the whole system can operate at 2 MHz. This frequency is not the upper bound of the 4F system since SLM could potentially operate in the GHz range, but in that case I/O might be the bottleneck of the system and complicates the assumption. Therefore GHz operating frequency is not assumed in this paper. 

\subsection{Leveraging Massive Optical Parallelism Using Computation Tiling}
\begin{table}[H]
\caption{Table of common notations used in this paper.}
\centering
\begin{tabularx}{\linewidth}{@{}cY@{}}
\toprule
Description                                 & Notation \\ \hline
Input size               & M $\times$ M           \\ 
Filter  size             & N $\times$ N           \\ 
SLM  size                & D $\times$ D           \\ 
Total number of filters                     & $N_{k}$     \\ 
Total number of input channels              & $N_{c}$     \\ 
Total number of inputs                      & $N_{i}$     \\ 
Number of blocks can be tiled on SLM & T           \\ \bottomrule
\end{tabularx}
\label{table:notation}
\end{table}

The light modulators for free-space 4F systems usually have high resolution. Phase mask can theoretically be fabricated to any resolution and high-end commercial SLM devices have resolution up to 4K \cite{4KDMD}. Take 4K SLM as an example, there are 16M pixels available to represent inputs or filter weights, which is much larger than almost any input/filter size used in CNNs. If only one input is convolved with one filter at a time, the SLMs (or phase masks) are severely under-utilized. Therefore a common approach to optimize the SLM utilization is tiling the inputs and/or filters across the SLM. 

\begin{figure}[h]
    \centering
    \includegraphics[width=8.5cm]{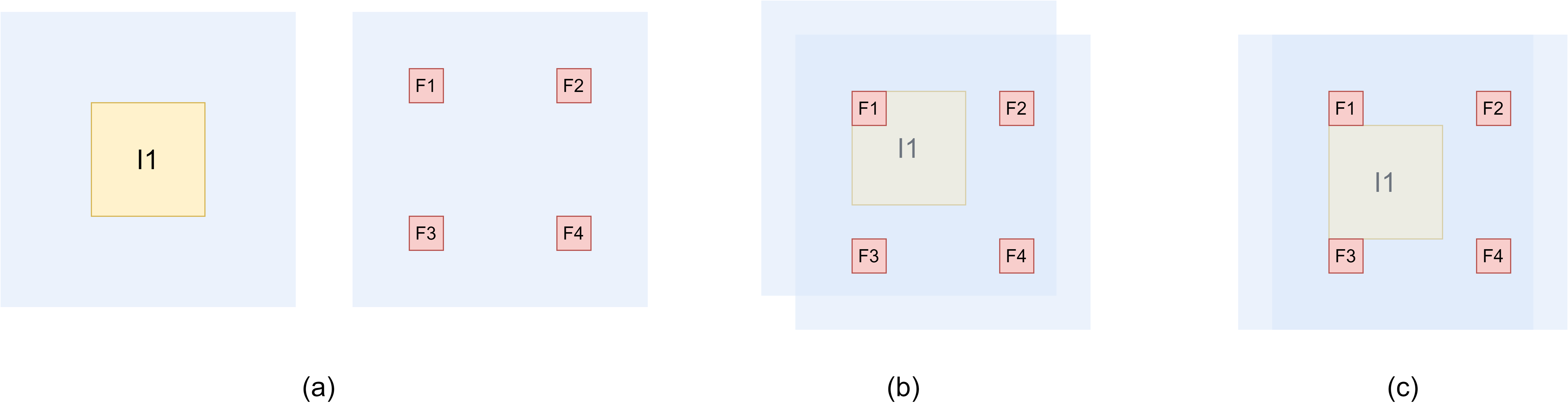}
    \caption{Illustration of filter tiling method. (a): Padded input block and tiled filter block. (b):Convolution visualization of filter f1 with the input. (c): Effect of zero padding. The padding between filters ensures that the input is not overlapped with two filters at same time. }
    \label{fig:kerneltiling}
\end{figure}

\cite{standford4f} adopts filter tiling to fully utilize the phase mask. For standard filter tiling, the filters are tiled in the space domain and then the Fourier transform of the tiled filter is loaded onto SLM or phase mask to perform the point-wise multiplication. The tiled filters convolve with only one input, similar to broadcasting. Filters need to be zero-padded to $(M+N-1) \times (M+N-1)$ as shown in fig \ref{fig:kerneltiling}(a) to generate valid results. The filter blocks are then tiled over the SLM to form a single large block and convolve with zero-padded input. Fig \ref{fig:kerneltiling}(b)(c) illustrates the convolution process. The output captured by the camera contains the tiled convolution results of the input with all individual filters tiled on the filter SLM/phase mask, which need to be extracted in the digital domain. Input tiling can be implemented in a similar way, but in this case inputs are tiled and convolved with a single filter. The tiled input can be directly loaded onto input SLM, since the Fourier transform of the input is carried out by the Fourier lens. 

To demonstrate the 4F system's advantage of 'free Fourier transform and multiplication', we compare a 4F system's (4K SLMs and 4K camera) convolution performance against a GPU (Nvidia RTX-2080 Ti) implementation of convolution using the Nvidia CuDNN library and FP16 precision. The CuDNN library contains three types of convolution algorithms, GEMM based algorithms, Winograd based algorithms and FFT based algorithms, and the optimal algorithm is selected for the given kernel/input size. Table \ref{tab:cudnncompare} shows the comparison between the 4F system and CuDNN implementations for different input sizes in terms of single convolution time. Single convolution time is the average time to perform a single 2D convolution operation, calculated by $$T_{single} = T_{total}/(N_{i}\times N_{c}\times N_{k})$$ where $T_{total}$ is the batch time for a single layer. For CuDNN implementation, batch size, number of kernels and channels are selected for optimal performance (full GPU utilization). For 4F implementation, a SLM based system with 4K resolution and 2 MHz frequency is assumed, based on the analysis in section \ref{hardwareintro}. Input tiling is applied to maximize the utilization of the 4F system. Three filter sizes are evaluated, which are $3\times 3$ (commonly used in modern CNNs), $7 \times 7$ (less common but still used in many networks) and $M \times M$ (same as input size, where 4F system's advantage is maximized). 
\begin{table*}[t]
\caption{Comparison of single convolution time (in seconds) between 4F SLM and CuDNN implementations, taking into account the effect of tiling and parallelism.}
\centering
\begin{tabularx}{\linewidth}{@{}Y|Y|YY|YY|YY@{}}
\toprule
\multicolumn{2}{c|}{} & \multicolumn{2}{c|}{N=3} & \multicolumn{2}{c|}{N=7} &\multicolumn{2}{c}{N=M} \\ \hline
Input size & $N_{c}$,$N_{k}$   & CuDNN    & 4F  & CuDNN   & 4F  & CuDNN   & 4F \\ \hline 
M=32       & 256,256 & 5.49-10   & 3.47e-11 & 2.22e-9 & 4.37e-11  & 8.21e-9   & 1.22e-10 \\
M=64       & 256,256 & 2.02e-9    & 1.30e-10 & 5.47e-9 & 1.48e-10 & 4.25e-8    & 4.88e-10 \\
M=128      & 128,128 & 9.12e-9  & 5.20e-10   & 1.92e-8 & 5.56e-10  & 9.56e-6   & 1.95e-9  \\
M=256      & 128,128 & 3.62e-8    & 2.22e-9 & 7.41e-8 & 2.22e-9 & 3.24e-3    & 7.81e-9  \\
M=512      & 64,64   & 1.93e-7   & 1.02e-8 & 6.83e-7 & 1.02e-8 & 2.29e-1  & 3.13e-8  \\
M=1024     & 32,32   & 1.59e-6    & 5.56e-8 & 4.04e-6 &5.56e-8  & 6.02e0   & 1.25e-7 \\ \bottomrule
\end{tabularx} 
\label{tab:cudnncompare}
\end{table*}
Based on table \ref{tab:cudnncompare}, for all input and filter sizes, a 2 MHz 4F system outperforms CuDNN implementation. The gap is larger when the filter size goes up, which is expected since the 4F system has $\mathcal{O}(1)$ complexity for convolution. For 4F implementation, the difference in convolution time for different filter sizes is trivial, only affected by the padding size (if tiling is applied), suggesting that the 4F system has more advantage on networks with large filters. 

\subsection{The Positive-only Photodetection Challenge}
For any optic/photonic CNN hardware, including the 4F system, the convolution/computation output needs to be converted to the digital domain by photodetectors or cameras, which apply a square function on the results by measuring the intensity. Since the camera/photodetector readouts are all positive, the weights need to be all-positive to make the individual convolution result valid (can use a square root function to retrieve the original result).
While this non-linearity could be potentially used as an activation function of CNN, it cannot be utilized on current 4F systems. Published 4F CNN hardware \cite{slm4f,standford4f} cannot compute the convolution of input with a multi-channel filter in one pass hence they need to repeat in a channel-by-channel manner and accumulate the partial sums in electronics (or can only process a fixed number of channels), which essentially puts the non-linearity before the channel summation and invalidates the convolution results. While positive weights might have a moderate impact on simple convolution networks with a small number of input channels for each layer, it will make the network barely functional for modern large networks with hundreds of channels each layer. Clearly, just using positive weights is not a solution and better methods are required to make the 4F system functional. Existing works deal with this limitation either optically \cite{Princeton_mzi} or computationally \cite{standford4f}.

Detecting the phase information using dedicated optical hardware can remove the positive weight restriction. \cite{Princeton_mzi} uses balanced photodetectors with MZM (Mach-Zehnder modulator) to detect the intensity and phase for photonic neuromorphic networks, though this technique cannot use in conventional cameras and may require precise alignment. \cite{opticalrelu} designs an optical ReLU module for the free-space 4F system, using SLMs and custom-built circuits. Polarization interferometry is used with a reference beam to detect the sign of each pixel and then feed the result back to SLM to let corresponding light pass through or deflect away. This method may be used to purely detect the sign information of the outputs detected by the camera for a conventional 4F system, at the cost of extra specialized hardware (at least double the number of photodetectors required and extra reference beam) and the challenge of integrating this method with conventional high-speed cameras (original design is based on SLMs).


\cite{standford4f} introduced a 'Pseudo-negative' method which can address the positive weight restriction without additional hardware, at the cost of doubling the number of filters and consequent computation overheads. The main idea of this method is to use all positive weights for each filter to ensure valid results, but label half of the filters as positive and the other half as negative. After the sensor readout, the results of negative filters are subtracted from the results of positive filters to form the final convolution results. The advantage of this approach is the convolution results can be considered identical to normal convolution due to the linearity of convolution (by using positive weights and taking square root after readout, the sensor non-linearity can be removed). However, the number of filters required is doubled, which means 2X weight memory and 0.5X throughput to replicate the original network.

\section{Channel Tiling for Optical Compute Parallelism}
\label{section:tiling}
Existing tiling approaches in optical computation overlook the fact that nearly all neural network layers in modern CNNs have multiple input channels giving another axis along which to parallelize the computation. To address the all-positive readout challenge, as well as eliminating the performance gap between SLM and high-speed cameras, we propose channel tiling (and mixed tiling) that can sum all channels inherently in the optical domain and with significantly lower output resolution requirement compared to other tiling methods. The channel tiling method tiles both input and filter channels and produces a single output, with convolution results for all channels accumulated (essentially implementing the multi-channel convolution optically). 

All tiling approaches are based on the assumption that the 4F system has the ability to multiply the inputs with complex-valued weights in the Fourier domain, which can be achieved by the system introduced in \ref{hardwareintro}. Thus, any real-valued filter can be transformed into the Fourier domain and multiplied with inputs without loss. The 4F system can then essentially be thought of as a black box convolution engine that can take inputs and filters each of size up to its resolution. In the analysis we assume SLM with resolution $D \times D$ is used for implementing point-wise multiplication in the Fourier domain, but the method also generalizes to other devices/components.

Tiling is done in the space domain and based on cross-correlation, which is more commonly used in the field of deep learning. To apply on a 4F system that performs convolution, filters need to be flipped before doing Fourier Transform. There are two common convolution modes ('same' and 'valid'), and they require slightly different tiling setups. For the 'same' mode where output has the same size as input, padding is essential during tiling to generate correct results. For 'valid' mode where the size of the output is $(M-N+1) \times (M-N+1)$ (see table \ref{table:notation}), padding is not necessary since the correct result can be extracted by down-sampling the output. In this section, all analysis is based on the 'same' mode which includes zero padding, while the tiling approach still holds for 'valid' mode except that padding is not required. For simplicity of analysis, actual SLM resolution is not taken into account for all tiling schemes except for mixed tiling. We validated all our analytical models of tiling (presented below) with computational experiments in a Python+Scipy+Numpy setup to confirm that tiling has no impact on the correctness of the convolution results.

\subsection{Channel Tiling Operation}
\begin{figure}[h]
    \centering
    \includegraphics[width=\linewidth]{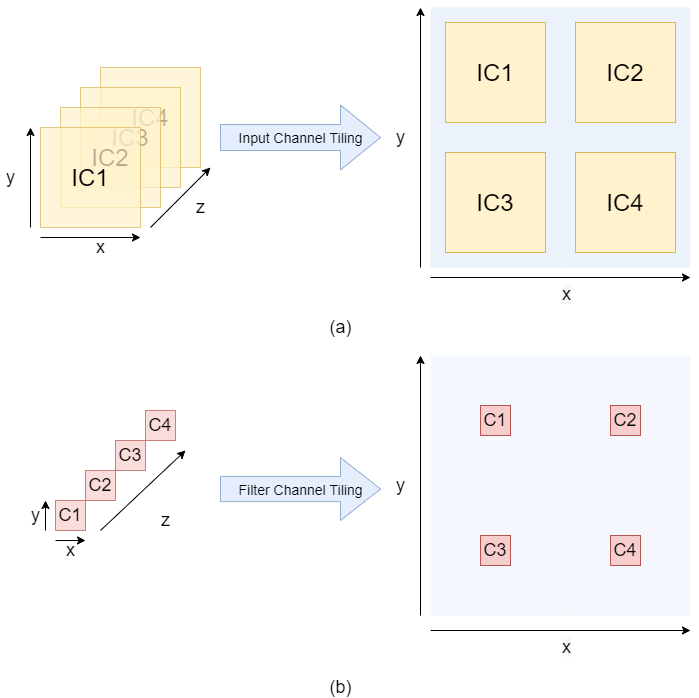}
    \caption{Illustration of tiling process of channel tiling, which effectively unrolls the channel dimension. Blue regions represents zero padding. (a): Input channel tiling. (b): Filter channel tiling. }
    \label{fig:channeltiling_tile}
\end{figure}

Consider the case where $N_{c}$ input channels with size $M \times M$ convolve with $N_{c}$ corresponding filters with size $N \times N$. The normal convolution process (same mode) can be formulated by 
\begin{equation}
    Y(i,j) = \sum_{a=0}^{N}\sum_{b=0}^{N} \sum_{c=0}^{N_{c}} X(a+i, b+j, c) \times F(a,b,c)
    \label{eqn:normalconv}
\end{equation}
where $X$ is zero padded input and $F$ is the convolution filter.
Channel tiling method tiles input and filter channels on corresponding 2D plane thus the summation over channels in the above formula is removed. Like other schemes, the channels for both input and filter need to be zero-padded into blocks with size $(M+N-1) \times (M+N-1)$, to avoid overlap between single filter channel with multiple input channels. Then both the input channel blocks and filter channel blocks are tiled in same order to form two large blocks $X_{T}$ and  $F_{T}$ with size $\ceil*{\sqrt{N_{c}}}(M+N-1) \times \ceil*{\sqrt{N_{c}}}(M+N-1)$, as shown in fig \ref{fig:channeltiling_tile}. For simplicity, denote the size of tiled input and filter blocks as $M_{t} \times M_{t}$, where $M_{t} = \ceil*{\sqrt{N_{c}}}(M+N-1)$. The formula for each output activation can be formulated by 
\begin{equation}
Y_{T}(i,j)=\sum_{a=0}^{M_{t}-1} \sum_{b=0}^{M_{t}-1} X_{TP}(a+i, b+j) \times F_{T}(a, b)
\end{equation}
where $X_{TP}$ is tiled input block $X_{T}$ and circular padded to size $(2M_{t}-1) \times (2M_{t}-1)$, which is inherently generated by Fourier transform. $F_{T}$ is the tiled filter block. The size of $Y_{T}$ and $F_{T}$ is $M_{t} \times M_{t}$. 

\begin{figure}[h]
    \centering
    \includegraphics[width=\linewidth]{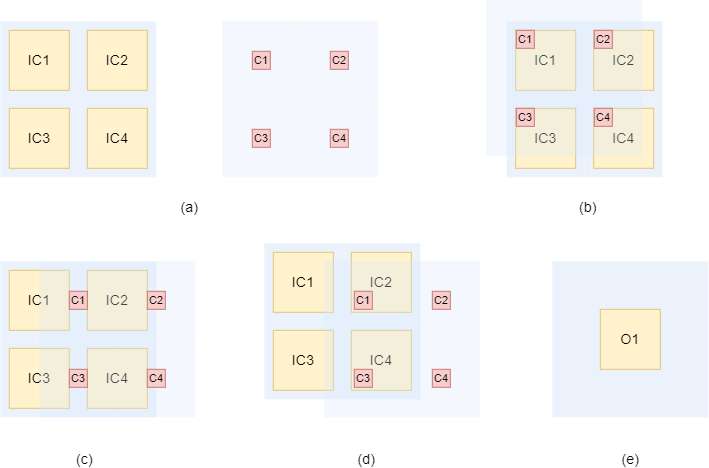}
    \caption{Illustration of the convolution process, blue regions represent zero padding. (a): Tiled input and filter blocks. (b): Start point of the valid region. (c): Effect of padding, filters will not overlap with multiple input channels. (d): Example of the invalid region, filters are not convolved with their corresponding input channels. (e): Output format of channel tiling.}
    \label{fig:channeltiling}
\end{figure}

When $i,j$ are in range of $ \frac{M_{t}-M}{2},\frac{M_{t}-M}{2}$ to $\frac{M_{t}-M}{2}+M,\frac{M_{t}-M}{2}+M $, the convolution result in this region can be expressed as

\begin{equation}
Y_{Tvalid}(i,j)=\sum_{a=0}^{M-1} \sum_{b=0}^{M-1} X_{T}(a+i, b+j) \times F_{T}(a, b)
\label{eqn:validregion}
\end{equation}
where $X_{T}$ is the original tiled input block before circular padding. Since input and filter channels are tiled in the same order, each individual input channel on the tiled input block aligns with its corresponding filter on the tiled filter block in this region, as shown in Fig. \ref{fig:channeltiling} (b). The sum over channel dimension in equation \ref{eqn:normalconv} is effectively unrolled into the other two dimensions thus equation \ref{eqn:validregion} has same output as equation \ref{eqn:normalconv}. Thus within this region the convolution result $Y_{T}$ is the standard multi-channel convolution result. 

When $i,j$ are outside of this region, as the case in Fig. \ref{fig:channeltiling} (c), the result is not valid, e.g., input channels are convolved wrong filter channels. Those invalid results make the $(M_{t}-1)/2$ extra zero padding of tiled input and filter due to inherent circular padding of Fourier transform unnecessary since circular padding only corrupts the results of invalid region. Therefore in this scheme only the center $M \times M$ region of the whole $M_{t} \times M_{t}$ convolution result is equivalent to the multi-channel convolution result of all input channels and should be extracted as the final result.

The output of this tiling scheme $Y_{T}$ is a single block with size $\ceil*{\sqrt{N_{c}}}(M+N-1)\ceil*{\sqrt{N_{c}}}(M+N-1)$ and only the center $M \times M$ region is valid. We can extract the valid region by directly selecting $y(i,j)$in range of $ \frac{M_{t}-M}{2},\frac{M_{t}-M}{2}$ to $\frac{M_{t}-M}{2}+M,\frac{M_{t}-M}{2}+M$. Fig \ref{fig:channeltiling}(e) visualizes the output format. Since only the center $M \times M$ part is used as the output, the camera's resolution requirement is massively reduced to the size of a single input. This property can significantly reduce output bandwidth and improve camera readout time.

We propose this novel channel tiling scheme to carry out the channel summation inherently in the optical domain, so that it won't be affected by the camera's non-linearity. By doing so the space domain filter values no longer need to be positive only, thus the 4F system can be modeled using absolute value or square function as an activation function with unconstrained filters during training, which is not possible in other tiling schemes.

\subsection{Mixed Tiling}
Since free-space 4F systems can support high resolution up to 4K, a large number of blocks can be tiled for inputs with small sizes. For some neural network structures, SLMs/phase masks cannot be fully tiled hence causing under-utilization. Taking Alexnet \cite{alexnet} as an example, for systems that support resolution higher than 1K, the system is far from fully utilized for any layer except the first one if channels or filters are tiled alone. To address this issue, we propose a mixed tiling scheme that combines channel tiling and filter tiling to improve the SLM utilization while still preserving the ability of channel tiling to carry out channel summation inherently in the optical domain.

Mixed tiling scheme is essentially applying two tiling schemes sequentially. The first step is applying channel tiling to tile channels of inputs and filters across a larger block. The tiled filter blocks are then further tiled across the filter SLM using filter tiling method, but zero-padding of individual tiled filter blocks is not necessary. The output is the convolution result of a multi-channel input against multiple filters with all channel results accumulated. By doing so the SLM utilization can be vastly improved compares with channel tiling scheme while channel summation capability is not affected. 

\begin{figure}[h]
    \centering
    \includegraphics[width=\linewidth]{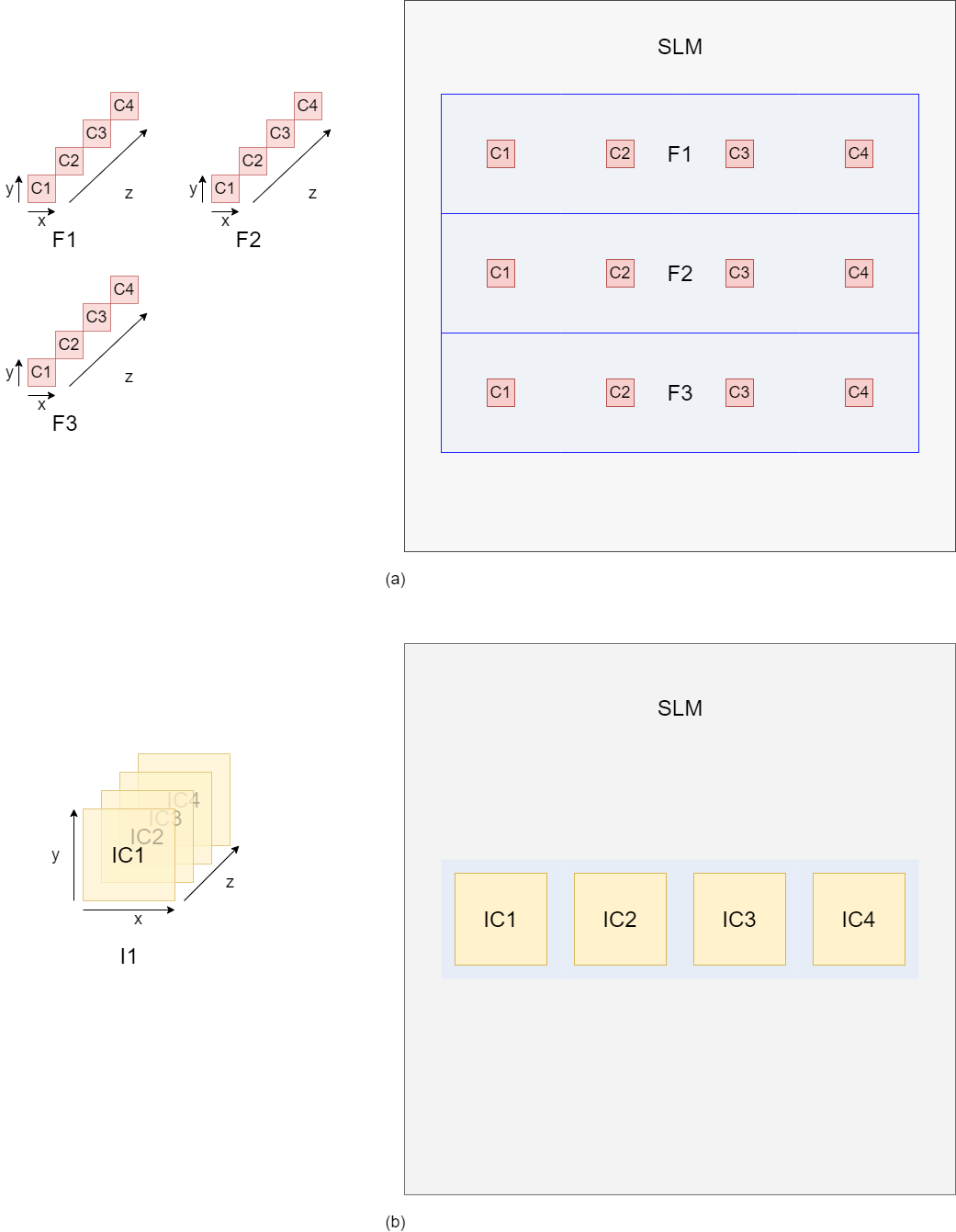}
    \caption{Illustration of tiling process of mixed tiling. Blue regions represents zero padding. (a): Filter channel tiling, multiple filters' channels are tiled on SLM. (b): Input channel tiling, only a single input's channels are tilied on SLM, follow the same order as filter channels.}
    \label{fig:mixedtiling}
\end{figure}

To maximize tiling efficiency, in step one, channels should be tiled horizontally into rows to better utilize the SLM resolution. 
The size of padded individual channel blocks are $(M+N-1)\times (M+N-1)$, the maximum number of such blocks that can be tiled on a single SLM is denoted as $T$, which equals to $\floor{\frac{D}{M+N-1}}^2$. The tiled block $B$ has shape $D,(M+N-1)\times\ceil{\frac{N_{c}}{\sqrt{T}}}$. The condition for mixed tiling is $N_{c} < \frac{T}{2}$, where $N_{c}$ is the total number of blocks that needs to be tiled for the first tiling scheme. 
\newline If the above condition is true, $B$ can be further tiled across all available SLM area from top to bottom. Figure \ref{fig:mixedtiling} (a) visualizes the filter plane where three filters and their channels are tiled and figure \ref{fig:mixedtiling} (b) visualizes the input plane where one input's channels are tiled. Zero padding inside $B$ blocks is not necessary, the padding within each single $(M+N-1)\times (M+N-1)$ block is enough to generate correct result, since only the invalid region (same as channel tiling) will be corrupted by overlapping of multiple filters.

The output format is a combination of the two applied tiling schemes. To extract the valid outputs, first split the raw output into results of individual filters and then extract the valid region inside each filter's result. Compared with other tiling schemes, the number of pixels required for detection is still low due to the combined channel tiling. However, since for mixed tiling the valid outputs are located at different regions, modification of the detection device is necessary to utilize the low output pixel count property in the mixed tiling case, such as a specialized controllable photodetector array. Without this, mixed tiling still delivers the required output bandwidth reduction from the camera but without reduction in resolution requirements.

\subsection{Tiling Efficiency Analysis and Optimization}
The tiling efficiency, or how much the SLM is utilized, is determined by several factors including the number of inputs/channels/filters, their sizes and whether padding is required or not. 

For the case where only a single tiling scheme is used, the number of blocks that needs to be tiled $N_{i,c,k}$ should be larger than $T$. For best performance, $N_{i,c,f}$ should be multiples of $T$ so that for each SLM update all pixels are fully utilized. The utilization rate is $$U = \frac{M^2\times N_{i,c,f}}{D^2\times \ceil{\frac{N_{i,c,f}}{T}}}$$ The numerator in equation is the total number of pixels for all blocks that require tiling and the denominator is total number of pixels actually used, which is a multiple of $D^2$. When $N_{i,c,f}$ is small compares with $T$, mixed tiling scheme should be adopted for optimal performance. 

For mixed tiling scheme, two tiling schemes are combined to improve SLM utilization. Again we use $B$ to denote the tiled blocks for the first scheme, $N_{1}$ and $N_{2}$ to denote the number of blocks that needs to be tiled for the first and second tiling scheme respectively (i.e., number of channels and filters). Since we assume SLMs are square shaped, the number of individual blocks that can be tiled in one row or column is $\sqrt{T}$. Similarly, the utilization of mixed tiling is 
$$U = \frac{M^2\times N_{1}\times N_{2}}{D^2\times \ceil{\frac{N_{2}}{T_{B}}}}$$
$T_{B}$, the total number of tiled block $B$ can be tiled across a single SLM, calculated by $$T_{B} = \floor[\Big]{\frac{\sqrt{T}}{\ceil{\frac{N_{1}}{\sqrt{T}}}}}$$

\subsection{Hardware Requirement Analysis}
Table \ref{tab:resolutiontable} shows the resolution requirements for different tiling schemes. Input and filter tiling are analogous to broadcasting where the untiled plane is broadcast to the tiled plane. For filter tiling (used in \cite{standford4f} along with pseudo-negative method), the input plane is not tiled therefore the input plane resolution is same as a single input. For input tiling, the filter plane is not tiled, but it still requires full resolution since filter plane is in the Fourier domain and must have the same resolution as input to perform point-wise multiplication. Channel-tiling and mixed tiling require full resolution of input and filter plane, while having very low resolution requirement for output. Modern CNN models usually have input image resolution lower than $300\times 300$ and are further downsampled in subsequent layers. Consider a 4K system setup with $300\times 300$ input resolution, {\em channel tiling will reduce the output resolution requirement by 186 times} compared with input/filter tiling when fully tiled. Such a massive reduction in output resolution will eliminate the performance gap between the output sensing unit and input/filter SLM. This reduction holds even for mixed tiling whose output resolution requirement is $N_{c}$ times less than input/filter tiling. $N_{c}$ is the number of input channels in a layer and is usually between 32 to 512 depends on exact network structure and layer. 
\begin{table}[h]
\caption{Comparison of the resolution requirements for different tiling schemes, assuming all cases are fully tiled.}
\centering
\begin{tabularx}{\linewidth}{@{}Y|Y|Y|Y@{}}
\toprule
Tiling schemes & Input res. & Filter res. & Output res. \\ \hline
None           & $M^2$            & $M^2$             & $M^2$             \\ 
Input tiling   & $D^2$            & $D^2$             & $D^2$             \\ 
Filter tiling \cite{standford4f}  & $M^2$            & $D^2$             & $D^2$    \\ 
Channel tiling & $D^2$            & $D^2$             & $M^2$             \\ 
Mixed tiling   & $D^2$            & $D^2$             & $\frac{D^2}{N_{c}}$    \\ \bottomrule
\end{tabularx}

\label{tab:resolutiontable}
\end{table}

\subsection{Impact of Camera Bit-Depth and Sensing Noise}
For the free-space 4F system, a high-speed camera is usually used as the output detection device and it adds two kinds of errors to the system, namely the quantization error and random sensing noise. Getting high bit-depth (i.e., precision), high resolution, high speed and low noise is a tough challenge for cameras and photodetectors. Most high speed cameras with reasonable resolution are limited to 8 or 12 bit precision (e.g., see \cite{phantom4kcams990,phantom4kcam}).

Due to limited precision or bit-depth of cameras, all outputs need to be quantized to 8-bit (or 12-bit) fixed-point format.  While conventional CNNs usually do not require high precision for inference and the impact of activation quantization on accuracy may be small, the case is different for optical systems since the square function is applied to the activation during sensing which increases the dynamic range and leading to larger quantization errors. The input and filter tiling quantize {\em each} channel as channel summation happens electronically after optical sensing (i.e., the partial sums themselves are quantized, not just the final activation value) thus quantization errors are propagated during the accumulation. In contrast, channel accumulation is carried out in the optical domain at full precision for the channel/mixed tiling and only the accumulated results are quantized to 8-bit or 12-bit, leading to a smaller overall quantization error. 

Similarly, channel tiling is less susceptible to sensing noise in the camera. Sensing noise can be especially limiting for fast, high-resolution cameras needed for optical computing. Random sensing noise increases error in every channel in input/filter tiling unlike channel tiling. The error scales with the number of channels thus it impacts more for larger networks. Furthermore, camera SNR (Signal to Noise ratio) scales with the photon flux (or number of photons captured by a pixel) \cite{camsnrconcept1,camsnrconcept2}. Intuitively, if the physical size of camera's sensor is fixed, then the higher resolution it supports (more pixels), the fewer photons each individual pixel will receive. As discussed, channel tiling (and mixed tiling) requires significantly less camera resolution compares to other tiling methods, which means it can have higher camera SNR than other methods. 

Channel tiling inference accuracy is expected to suffer much less from camera quantization and noise. Results illustrating this benefit of channel tiling are discussed in Section \ref{sec:camerasnr}.

\section{Evaluation and Results}
In this section, we evaluate and compare the proposed channel/mixed tiling approaches with other state-of-the-art 4F optics approaches, both in terms of performance and inference accuracy. 

\subsection{Comparing Performance of Tiling Approaches} 
To compare the tiling efficiency of different tiling methods, we compare them against a Nvidia RTX-2080 Ti GPU with FP16 precision using inference time on real networks. Inference time is the average time for a single input to pass through the whole network. For GPU, the batch size is set for full utilization while for 4F system a 4K, 2 MHz SLM based system is assumed. Only convolution layers are benchmarked. We pick VGG16 \cite{vgg16} (with CIFAR, ImageNet and SpaceNet \cite{van2018spacenet}) and AlexNet \cite{alexnet} (with ImageNet) as representative benchmarks. Besides these two widely used CNN architectures for image classification, we also include two other networks for image super-resolution and de-convolution. For image super-resolution, the SRCNN \cite{srcnn} is used, which consists of two convolution layers with filter size $9\times 9$ and $5\times 5$ and the target image resolution is set to $512 \times 512$. The Deconv Net \cite{NIPS2014_5485} is used for image de-convolution and contains five convolution layers with large filter sizes ($121\times 1, 1\times 121, 16\times 16, 1\times 1$, and $8\times 8$ respectively). 

The 2MHz SLM system is faster than GPU in all cases and is as much as 61.7X faster for the SRCNN case, which is better suited to the 4F system given its larger kernel sizes. For most conventional neural network architectures with relatively small filter sizes (e.g., AlexNet and VGG), the speedup is around 20X. It is interesting to compare VGG16 performance on CIFAR10 vs. ImageNet. The smaller input size ($32\times 32$ vs. $227\times 227$) of CIFAR10 dataset leads to severe underutilization of the 4K SLM, especially in later layers of the network. This indicates that smaller (and therefore potentially cheaper, faster SLMs) may be a better design point for small input networks. 

Since the main goal of this paper is to focus on bringing 4F optical computing closer to reality, the complete analysis of energy and power is out of scope of this paper. The analysis of power and energy of various photonic and optical neural network implementations can be found in \cite{photonicnnsurvvey}.

\begin{table*}[t]
\caption{Overall network inference time in seconds (per input) for different tiling schemes and network architectures. The results are for convolution layers only. Note for DeconvNet there's a convolution layer of 1x1 filters which is not suitable for 4F system acceleration and is not taken into account for the estimation.}
\centering
\begin{tabularx}{\linewidth}{@{}Y|Y|Y|Y|Y|Y@{}}
\toprule
Network-dataset  & GPU    & Channel tiling & Mixed tiling & No Tiling    &Best speedup     \\ \hline
VGG16-CIFAR10    & 5.07e-5  & 1.98e-3 & 6e-6   & 8.17e-1   &\textbf{8.45}  \\ 
VGG16-ImageNet  & 1.41e-3  & 1.98e-3   & 6e-5   & 8.17e-1   &\textbf{23.50}\\ 
AlexNet-ImageNet & 1.31e-4  & 6.88e-4  & 7e-6   & 1.84e-1   &\textbf{18.71}   \\ 
VGG16-SpaceNet7 & 1.79e-2   &2.27e-3    & 1.06e-3   &8.17e-1 &\textbf{16.89} \\ 
DeconvNet     & 3.76e-4  & 3.35e-4      &8e-6    &1.02e-2 &\textbf{47.00} \\ 
SRCNN       & 1.48e-3 & 9.60e-5 & 2.4e-5     & 2.88e-4  &  \textbf{61.67} \\ \bottomrule
\end{tabularx}
\label{tab:networkcomp}
\end{table*}

\subsection{Impact of Tiling Approaches on Network Accuracy}
As discussed previously, different tiling schemes impose different restrictions on the network: input/filter tiling places a square function on each channel's convolution result before summation, while channel/mixed tiling restricts the activation function to be absolute value function (taking a square root after camera readout). The effect of these restrictions on network-level accuracy is reported in table \ref{tab:accuracycomp}, evaluating three datasets trained with a VGG-16 like model. The camera is assumed to have unlimited precision (addressed in the next section). All-positive filters are not used in input/filter tiling methods as it effectively nullifies the purpose of activation and makes deep CNNs like VGG-16 extremely hard to train properly since there are (almost always) multiple input channels. Optical compensation approaches (e.g., the sign detection proposed in \cite{opticalrelu}) and pseudo-negative \cite{standford4f}, in the absence of any optical non-idealities, should give the same accuracy as standard convolution. For the proposed channel/mixed tiling, the network is trained with absolute value function applied to convolution results. All the results reported in table \ref{tab:accuracycomp} are trained from scratch with floating point precision.

The results clearly show that input or filter tiling is unacceptably inaccurate for anything but the simplest of classification tasks. Our proposed channel and mixed tiling approaches lose $<$3 percent accuracy compares to unconstrained electronic implementations or pseudo-negative approach (which uses twice as many filters). This small gap in accuracy can be bridged in future by better optical non-linearities (an active area of research \cite{nonlinearity_1,nonlinearity_2}), improved training methods (e.g. different regularization terms during training to incentivize positive filter weights) or combining with the pseudo-negative approach \cite{standford4f}, all are part of our ongoing work but not explored further in this paper. 

\begin{table}[h]
\caption{Comparison of the accuracy of different datasets trained using VGG16 with different methods. For Input and filter tiling the absolute value function is applied to each individual channel's convolution result while for channel and mixed tiling the absolute value function is applied after channel summation and acts as the activation function. For pseudo-negative case, positive weights are used and subtraction is implemented after camera detection, then ReLU is applied on subtracted results as the activation function.}
\begin{tabularx}{\linewidth}{@{}Y|Y|Y|Y@{}}
\toprule
Method                      & Fashion MNIST & SVHN   & CIFAR10 \\ \hline
Input/Filter Tiling   & 75.4\%       &  78.5\%      & 55.6\%  \\ 
\textbf{Channel/mixed Tiling}  & 93.2\%       & 95.1\% & 89.3\%  \\ 
Pseudo negative    & 93.6\%    & 95.1\% & 91.6\%  \\ \bottomrule
\end{tabularx}
\label{tab:accuracycomp}
\end{table}

\subsection{Impact of Camera Limitations on Inference Accuracy}\label{sec:camerasnr}
The previous accuracy results are ideal cases and do not consider camera bit-precision and sensing noise. 
To simulate quantization error due to the camera's limited bit-precision, square function is applied to the convolution results for simulating the intensity measurement and then the results are then quantized to scaled 8-bit and 12-bit format (256/4096 uniform intervals). The square root is taken on the quantized results to get the absolute value. We simulate the sensing noise using white Gaussian noise with average SNR at 15dB, 20dB and 30dB \cite{snrresult}. For both cases only partial sum and activations are quantized, while weights as assumed to be floating point precision. \footnote{We do not use high SNR values as high resolution, high-speed cameras are likely to have higher photon noise \cite{snrresult}.}

\begin{figure}[h]
    \centering
    \includegraphics[width=\linewidth]{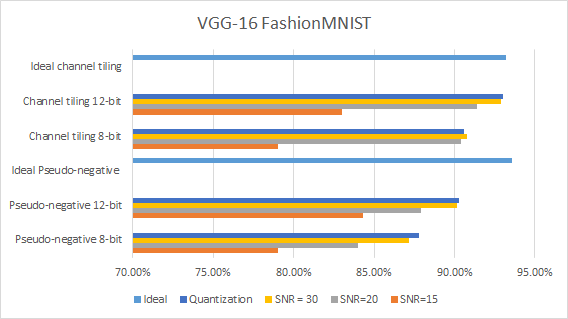}
    \caption{Accuracy of FashionMNIST dataset using different setups.}
    \label{fig:barresult_fashion}
\end{figure}

\begin{figure}[h]
    \centering
    \includegraphics[width=\linewidth]{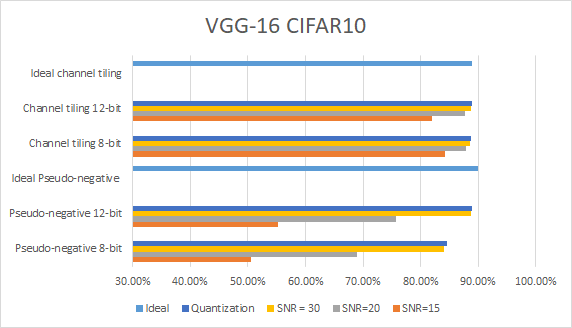}
    \caption{Accuracy of CIFAR-10 dataset using different setups.}
    \label{fig:barresult_cifar}
\end{figure}

Figure \ref{fig:barresult_fashion} and figure \ref{fig:barresult_cifar} show the Fashion MNIST and CIFAR-10 classification accuracy using VGG-16 for pseudo-negative with filter tiling and the proposed channel/mixed tiling, taking into account camera quantization error and different level of random noise. The results clearly show that channel/mixed tiling is far more robust to both quantization and sensing noise due to its error-free channel summation. The pseudo-negative (filter tiling) approach requires at least 12-bit camera precision as opposed to 8-bit for channel/mixed tiling to remain within \emph{5 percent} accuracy drop from full precision. Interestingly, once the camera bit-depth is taken into account, channel tiling always has higher accuracy than a pseudo-negative approach despite being somewhat more restrictive. 
 
For cases with random noise, the accuracy of the proposed channel/mixed tiling method is higher than the pseudo-negative method for almost all cases. 30dB or 20dB SNR is good enough for channel tiling accuracy to be within \emph{5 percent} of the noiseless case while other tiling approaches need at least 30dB SNR. Moreover, the achievable SNR for channel/mixed tiling can be higher than filter tiling since it requires lower output resolution, making channel/mixed tiling more robust to sensing noise.

Altogether, our results indicate that the proposed channel tiling approach can reduce required camera precision by \emph{33 percent} (8-bit vs. 12-bit) and improve noise tolerance by \emph{10dB}. Though we do not explore it here, such relaxation can substantially improve the speed, energy, and cost of sensing in 4F computing systems.

\section{Discussion}

\subsection{Experimental Realization of Programmable 4F System}
\begin{figure}[h]
    \centering
    \includegraphics[width=\linewidth]{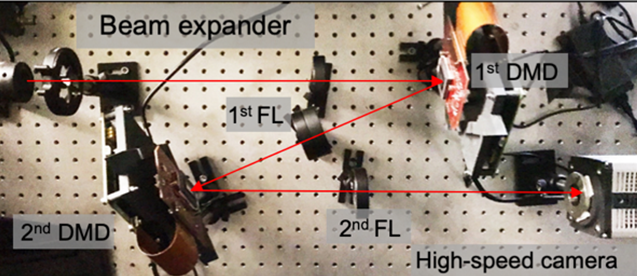}
    \caption{Experimental setup of the DMD based programmable 4F system.}
    \label{fig:expsetup}
\end{figure}

One primary assumption in this paper is an SLM based, programmable 4F system. The experimental verification is not the focus of this paper, and the realization of such system has been demonstrated by our recent work \cite{miscuglio2020optica}. In that work a prototype of a DMD based high-speed programmable 4F system has been built with KHz operating frequency, and able to get decent inference accuracy for a single layer CNN. Figure \ref{fig:expsetup} shows the photo of experimental setup of the programmable 4F system. Unlike passive 4f systems where filters are fixed, an active 4F system requires filter weights to feed into the DMD at high speed consistently, thus high-speed I/O and synchronization between SLM and camera become challenges to the system. An FPGA based unified high-speed I/O interface is designed to manage the I/O and control of the whole 4F system. Besides, a training-compatible simulation model is designed to accurately model the 4F system during network training while fine-tuning is applied to compensate various system non-idealities, addressing two other common issues of 4F systems.

This experimental realization of a programmable 4F system presents a way to scale the 4F system and make it target a broader aspect of applications. As discussed in section \ref{hardwareintro}, the SLM modulation speed is continuously improving with new techniques and materials, and the trend is highly encouraging: although commercial available SLM only operates in KHz range \cite{1080p23kDMD,4KDMD}, GHz SLM \cite{Peng19ghzslm} is with-in reach with advanced materials. The advancement of SLM technology makes the 4F computing system promising, as it scales better than conventional electronic computing systems.

\subsection{Optimal Filter Size and Architecture Search}
One major trend in CNN architecture is making the filter size (pixel size) as small as possible, $5 \times 5$ and larger filters are rarely seen in recent years. This trend is fully understandable, as almost all current CNNs need to be trained and implemented on electronic systems, whose computation time scales with filter size. Using a small filter size with a deeper structure will make the neural network a lot faster and still have a similar receptive field size. Following this trend, many researchers developed various successful CNN architectures with very small filter sizes and making a small filter size become the de-facto choice for CNN architectures.

However, it is a completely different story for 4F systems. Based on the complexity analysis and experiment results from table \ref{tab:cudnncompare}, increasing filter size has a trivial impact on convolution speed for 4F systems, which indicates that larger filters may be more suitable for 4F system as its almost 'free'. Unfortunately, due to the dominance of electronics in the field of computer vision, the potential of large filters has not been thoroughly explored, due to this computational inefficiency. We argue that large filter sizes should be of interest in many use cases if the computation efficiency burden is removed. For example, the image resolution used in computer vision tasks is constantly growing: A smartphone photo can have 4K resolution and some other applications like medical images or satellite images have much higher resolution. If those high-resolution images are fed into a model designed for medium-resolution images, without proper scaling, a $3\times 3$ filter will capture way less information than intended and may not be the optimal filter size anymore. Thus, the advancement of 4F systems enables a new CNN architecture search direction: a shallower network with large filters. In this region, 4F systems could operate significantly more efficiently than electronic systems, and thus able to support architectures that are not possible using electronic systems due to performance reasons. 

\section{Conclusion}
In this work, we provide scalability analysis and performance estimation of 4F optical systems on CNN acceleration, along with related issues. We then propose the channel tiling and mixed tiling methods for optical 4F systems to boost the performance and accuracy, without extra hardware or computation. By utilizing the properties of convolution and 4F systems' high resolution, channel tiling and mixed tiling make 4F systems able to accumulate all channel's convolution results in the optical domain, thus bypassing various constraints applied by output sensing. Compared to the recent pseudo-negative approach with filter tiling \cite{standford4f}, our method gives similar accuracy ($<$3 percent difference on three datasets), significantly better robustness to sensing quantization error (33 percent improvement in required sensing precision) and noise (10dB reduction in tolerable sensing noise), 0.5X total filters required, 10-50X+ throughput improvement and at least 3X reduction in required output camera resolution/bandwidth. The proposed channel tiling and mixed tiling methods provide a simple and practical way to fully utilize the massive parallelism inherent in 4F optical computing systems to accelerate CNNs and bring it closer to reality.

\AtNextBibliography{\scriptsize}
\printbibliography

@incollection{NIPS2014_5485,
title = {Deep Convolutional Neural Network for Image Deconvolution},
author = {Xu, Li and Ren, Jimmy SJ and Liu, Ce and Jia, Jiaya},
booktitle = {Advances in Neural Information Processing Systems 27},
editor = {Z. Ghahramani and M. Welling and C. Cortes and N. D. Lawrence and K. Q. Weinberger},
pages = {1790--1798},
year = {2014},
publisher = {Curran Associates, Inc.},
url = {http://papers.nips.cc/paper/5485-deep-convolutional-neural-network-for-image-deconvolution.pdf}
}

@inproceedings{DMDconcept,
	author={Jeffrey B. Sampsell},
	year={Nov 1994},
	title={Digital micromirror device and its application to projection displays},
	booktitle = {Journal of Vacuum Science \& Technology B: Microelectronics and Nanometer Structures},
	volume={12},
	chapter={6},
	pages={3242-3246},
	isbn={0734-211X},
	language={English},
	url={http://dx.doi.org/10.1116/1.587506},
	doi={10.1116/1.587506}
}

@misc{4KDMD,
    author={{Texas Instruments}},
	title={DLP470TE datasheet},
	note={\url{https://www.ti.com/document-viewer/DLP470TE/datasheet/features-dlps1611610}}
}

@misc{1080p23kDMD,
    author={{Texas Instruments}},
	title={DLP9500 datasheet},
	note={\url{https://www.ti.com/document-viewer/DLP9500/datasheet}}
}

@inproceedings{IPM1mhzdmd,
author = {Jan-Uwe Schmidt and Ulrike A. Dauderstaedt and Peter Duerr and Martin Friedrichs and Thomas Hughes and Thomas Ludewig and Dirk Rudloff and Tino Schwaten and Daniela Trenkler and Michael Wagner and Ingo Wullinger and Andreas Bergstrom and Peter Bjoernangen and Fredrik Jonsson and Tord Karlin and Peter Ronnholm and Torbjorn Sandstrom},
title = {{High-speed one-dimensional spatial light modulator for Laser Direct Imaging and other patterning applications }},
volume = {8977},
booktitle = {MOEMS and Miniaturized Systems XIII},
editor = {Wibool Piyawattanametha and Yong-Hwa Park},
organization = {International Society for Optics and Photonics},
publisher = {SPIE},
pages = {167 -- 176},
year = {2014},
doi = {10.1117/12.2036533},
URL = {https://doi.org/10.1117/12.2036533}
}

@INPROCEEDINGS{2.3MHzDMD,

  author={L. {Haspeslagh} and J. {De Coster} and O. V. {Pedreira} and I. {De Wolf} and B. {Du Bois} and A. {Verbist} and R. {Van Hoof} and M. {Willegems} and S. {Locorotondo} and G. {Bryce} and J. {Vaes} and B. {van Drieenhuizen} and A. {Witvrouw}},

  booktitle={2008 IEEE International Electron Devices Meeting}, 

  title={Highly reliable CMOS-integrated 11MPixel SiGe-based micro-mirror arrays for high-end industrial applications}, 

  year={2008},

  volume={},

  number={},

  pages={1-4},}

@misc{phantom4kcam,
    author={{Phantom}},
	title={Flex4k Camera},
	note={\url{https://www.phantomhighspeed.com/products/cameras/4kmedia/flex4k}}
}

@misc{phantom4kcams990,
    author={{Phantom}},
	title={S990 Camera},
	note={\url{https://www.phantomhighspeed.com/products/cameras/machinevision/s990}}
}

@misc{SpecializedImaging7M,
    author={{Specialized Imaging}},
	title={Kirana7M Camera},
	note={\url{https://www.specialised-imaging.com/products/video-cameras/kirana}}
}

@misc{SpecializedImaging1B,
    author={{Specialized Imaging}},
	title={SIMX Camera},
	note={\url{https://www.specialised-imaging.com/products/framing-cameras/simx}}
}

@inproceedings{analogmmaref1,
author = {Hubert K. Lakner and Peter Duerr and Ulrike Dauderstaedt and Wolfgang Doleschal and Joerg Amelung},
title = {{Design and fabrication of micromirror arrays for UV lithography}},
volume = {4561},
booktitle = {MOEMS and Miniaturized Systems II},
editor = {M. Edward Motamedi and Rolf Goering},
organization = {International Society for Optics and Photonics},
publisher = {SPIE},
pages = {255 -- 264},
year = {2001},
doi = {10.1117/12.443094},
URL = {https://doi.org/10.1117/12.443094}
}

@misc{HOLOEYESLM,
    author={{HOLOEYE}},
	title={GAEA-2},
	note={\url{https://holoeye.com/gaea-4k-phase-only-spatial-light-modulator/}}
}

@misc{700kslm,
    author={{Meadowlark Optics}},
	title={1920slm},
	note={\url{https://www.meadowlark.com/1920-1152-spatial-light-modulator-p-119}}
}

@misc{400kslm,
    author={{AVR Optics}},
	title={HSP1920},
	note={\url{https://www.avr-optics.com/catalog/spatial_light_modulators___scanning_systems/xy_slm/hsp1920}}
}

@article{photonicnnsurvvey,
  title={Photonic Neural Networks: A Survey},
  author={De Marinis, Lorenzo and Cococcioni, Marco and Castoldi, Piero and Andriolli, Nicola},
  journal={IEEE Access},
  volume={7},
  pages={175827--175841},
  year={2019},
  publisher={IEEE}
}

@article{mitopticcnn,
  title={On-chip optical convolutional neural networks},
  author={Bagherian, Hengameh and Skirlo, Scott and Shen, Yichen and Meng, Huaiyu and Ceperic, Vladimir and Soljacic, Marin},
  journal={arXiv preprint arXiv:1808.03303},
  year={2018}
}

@article{mrr2019,
  title={Digital electronics and analog photonics for convolutional neural networks (DEAP-CNNs)},
  author={Bangari, Viraj and Marquez, Bicky A and Miller, Heidi and Tait, Alexander N and Nahmias, Mitchell A and de Lima, Thomas Ferreira and Peng, Hsuan-Tung and Prucnal, Paul R and Shastri, Bhavin J},
  journal={IEEE Journal of Selected Topics in Quantum Electronics},
  volume={26},
  number={1},
  pages={1--13},
  year={2019},
  publisher={IEEE}
}

@inproceedings{pcnna,
  title={PCNNA: a photonic convolutional neural network accelerator},
  author={Mehrabian, Armin and Al-Kabani, Yousra and Sorger, Volker J and El-Ghazawi, Tarek},
  booktitle={2018 31st IEEE International System-on-Chip Conference (SOCC)},
  pages={169--173},
  year={2018},
  organization={IEEE}
}

@article{slm4f,
  title={Optical frontend for a convolutional neural network},
  author={Colburn, Shane and Chu, Yi and Shilzerman, Eli and Majumdar, Arka},
  journal={Applied optics},
  volume={58},
  number={12},
  pages={3179--3186},
  year={2019},
  publisher={Optical Society of America}
}

@article{standford4f,
  title={Hybrid optical-electronic convolutional neural networks with optimized diffractive optics for image classification},
  author={Chang, Julie and Sitzmann, Vincent and Dun, Xiong and Heidrich, Wolfgang and Wetzstein, Gordon},
  journal={Scientific reports},
  volume={8},
  number={1},
  pages={1--10},
  year={2018},
  publisher={Nature Publishing Group}
}

@article{slmmodulatingref,
  title={Arbitrary manipulation of spatial amplitude and phase using phase-only spatial light modulators},
  author={Zhu, Long and Wang, Jian},
  journal={Scientific reports},
  volume={4},
  pages={7441},
  year={2014},
  publisher={Nature Publishing Group}
}

@article{oamDMD,
  title={Rapid generation of light beams carrying orbital angular momentum},
  author={Mirhosseini, Mohammad and Magana-Loaiza, Omar S and Chen, Changchen and Rodenburg, Brandon and Malik, Mehul and Boyd, Robert W},
  journal={Optics express},
  volume={21},
  number={25},
  pages={30196--30203},
  year={2013},
  publisher={Optical Society of America}
}

@article{dmdtheory,
  title={Binary synthetic holograms},
  author={Lee, Wai-Hon},
  journal={Applied optics},
  volume={13},
  number={7},
  pages={1677--1682},
  year={1974},
  publisher={Optical Society of America}
}

@article{superpixeldmdmodulation,
  title={Superpixel-based spatial amplitude and phase modulation using a digital micromirror device},
  author={Goorden, Sebastianus A and Bertolotti, Jacopo and Mosk, Allard P},
  journal={Optics express},
  volume={22},
  number={15},
  pages={17999--18009},
  year={2014},
  publisher={Optical Society of America}
}

@article{mziref,
  title={Ueber einen interferenzrefraktor},
  author={Mach, Ludwig},
  journal={Zeitschrift f{\"u}r Instrumentenkunde},
  volume={12},
  number={3},
  pages={89},
  year={1892}
}

@article{phaseencodedmd,
  title={High sampling rate single-pixel digital holography system employing a DMD and phase-encoded patterns},
  author={Gonz{\'a}lez, Humberto and Mart{\'\i}nez-Le{\'o}n, Llu{\'\i}s and Soldevila, Fernando and Araiza-Esquivel, Ma and Lancis, Jes{\'u}s and Tajahuerce, Enrique},
  journal={Optics express},
  volume={26},
  number={16},
  pages={20342--20350},
  year={2018},
  publisher={Optical Society of America}
}

@ARTICLE{11GSDAC,

  author={E. {Olieman} and A. {Annema} and B. {Nauta}},

  journal={IEEE Journal of Solid-State Circuits}, 

  title={An Interleaved Full Nyquist High-Speed DAC Technique}, 

  year={2015},

  volume={50},

  number={3},

  pages={704-713},}

@INPROCEEDINGS{12GSDAC,

  author={Q. {Ye} and Y. {Zhang} and X. {Li} and Y. {Zhang}},

  booktitle={2017 2nd IEEE International Conference on Integrated Circuits and Microsystems (ICICM)}, 

  title={A 12-bit 10GSps ultra high speed DAC in InP HBT technology}, 

  year={2017},

  volume={},

  number={},

  pages={9-13},}

@article{vgg16,
  title={Very deep convolutional networks for large-scale image recognition},
  author={Simonyan, Karen and Zisserman, Andrew},
  journal={arXiv preprint arXiv:1409.1556},
  year={2014}
}

@ARTICLE{nonlinearity_1,

  author={I. A. D. {Williamson} and T. W. {Hughes} and M. {Minkov} and B. {Bartlett} and S. {Pai} and S. {Fan}},

  journal={IEEE Journal of Selected Topics in Quantum Electronics}, 

  title={Reprogrammable Electro-Optic Nonlinear Activation Functions for Optical Neural Networks}, 

  year={2020},

  volume={26},

  number={1},

  pages={1-12},}

@inproceedings{nonlinearity_2,
author = {Jonathan George and Amin Mehrabian and Rubab Amin and Tarek El-Ghazawi and Paul K. Prucnal and Volker J. Sorger},
booktitle = {Frontiers in Optics / Laser Science},
journal = {Frontiers in Optics / Laser Science},
pages = {JW3A.123},
publisher = {Optical Society of America},
title = {Photonic Neural Network Nonlinear Activation Functions by Electrooptic Absorption Modulators},
year = {2018},
url = {http://www.osapublishing.org/abstract.cfm?URI=LS-2018-JW3A.123},
}

@inproceedings{alexnet,
  title={Imagenet classification with deep convolutional neural networks},
  author={Krizhevsky, Alex and Sutskever, Ilya and Hinton, Geoffrey E},
  booktitle={Advances in neural information processing systems},
  pages={1097--1105},
  year={2012}
}

@article{opticalrelu,
  title={Optical rectifying linear units for back-propagation learning in a deep holographic convolutional neural network},
  author={Wagner, Kelvin H and McComb, Sean},
  journal={IEEE Journal of Selected Topics in Quantum Electronics},
  volume={26},
  number={1},
  pages={1--18},
  year={2019},
  publisher={IEEE}
}

@article{Princeton_mzi,
  title={Neuromorphic photonic networks using silicon photonic weight banks},
  author={Tait, Alexander N and De Lima, Thomas Ferreira and Zhou, Ellen and Wu, Allie X and Nahmias, Mitchell A and Shastri, Bhavin J and Prucnal, Paul R},
  journal={Scientific reports},
  volume={7},
  number={1},
  pages={1--10},
  year={2017},
  publisher={Nature Publishing Group}
}

@INPROCEEDINGS {camsnrconcept1,
author = {G. Healey and R. Kondepudy},
booktitle = {Proceedings 1992 IEEE Computer Society Conference on Computer Vision and Pattern Recognition},
title = {CCD camera calibration and noise estimation},
year = {1992},
volume = {},
issn = {1063-6919},
pages = {90,91,92,93,94,95},
doi = {10.1109/CVPR.1992.223222},
url = {https://doi.ieeecomputersociety.org/10.1109/CVPR.1992.223222},
publisher = {IEEE Computer Society},
address = {Los Alamitos, CA, USA},
month = {jun}
}

@inproceedings{camsnrconcept2,
author = {Glenn Healey and Raghava V. Kondepudy},
title = {{Modeling and calibrating CCD cameras for illumination-insensitive machine vision}},
volume = {1614},
booktitle = {Optics, Illumination, and Image Sensing for Machine Vision VI},
editor = {Donald J. Svetkoff},
organization = {International Society for Optics and Photonics},
publisher = {SPIE},
pages = {121 -- 132},
year = {1992},
doi = {10.1117/12.57974},
URL = {https://doi.org/10.1117/12.57974}
}

@article{photonicmac_2019photonic,
  title={Photonic multiply-accumulate operations for neural networks},
  author={Nahmias, Mitchell A and De Lima, Thomas Ferreira and Tait, Alexander N and Peng, Hsuan-Tung and Shastri, Bhavin J and Prucnal, Paul R},
  journal={IEEE Journal of Selected Topics in Quantum Electronics},
  volume={26},
  number={1},
  pages={1--18},
  year={2019},
  publisher={IEEE}
}

@article{snrresult,
  title={Characteristics and performance of image sensor communication},
  author={Huang, Wei and Xu, Zhengyuan},
  journal={IEEE Photonics Journal},
  volume={9},
  number={2},
  pages={1--19},
  year={2017},
  publisher={IEEE}
}

@article{miscuglio2020optica,
author = {Mario Miscuglio and Zibo Hu and Shurui Li and Jonathan K. George and Roberto Capanna and Hamed Dalir and Philippe M. Bardet and Puneet Gupta and Volker J. Sorger},
journal = {Optica},
number = {12},
pages = {1812--1819},
publisher = {OSA},
title = {Massively parallel amplitude-only Fourier neural network},
volume = {7},
month = {Dec},
year = {2020},
url = {http://www.osapublishing.org/optica/abstract.cfm?URI=optica-7-12-1812},
doi = {10.1364/OPTICA.408659}
}

@article{Peng19ghzslm,
author = {Cheng Peng and Ryan Hamerly and Mohammad Soltani and Dirk R. Englund},
journal = {Opt. Express},
number = {21},
pages = {30669--30680},
publisher = {OSA},
title = {Design of high-speed phase-only spatial light modulators with two-dimensional tunable microcavity arrays},
volume = {27},
month = {Oct},
year = {2019},
url = {http://www.opticsexpress.org/abstract.cfm?URI=oe-27-21-30669},
doi = {10.1364/OE.27.030669},
}

@article{van2018spacenet,
  title={Spacenet: A remote sensing dataset and challenge series},
  author={Van Etten, Adam and Lindenbaum, Dave and Bacastow, Todd M},
  journal={arXiv preprint arXiv:1807.01232},
  year={2018}
}

@ARTICLE{srcnn,
  author={C. {Dong} and C. C. {Loy} and K. {He} and X. {Tang}},
  journal={IEEE Transactions on Pattern Analysis and Machine Intelligence}, 
  title={Image Super-Resolution Using Deep Convolutional Networks}, 
  year={2016},
  volume={38},
  number={2},
  pages={295-307},
  doi={10.1109/TPAMI.2015.2439281}}

@article{complexbinarydmd,
  title={Full-complex amplitude modulation with binary spatial light modulators},
  author={Ulusoy, Erdem and Onural, Levent and Ozaktas, Haldun M},
  journal={JOSA A},
  volume={28},
  number={11},
  pages={2310--2321},
  year={2011},
  publisher={Optical Society of America}
}

@article{arbitrarycomplexusingphaseslmnature,
  title={Arbitrary manipulation of spatial amplitude and phase using phase-only spatial light modulators},
  author={Zhu, Long and Wang, Jian},
  journal={Scientific reports},
  volume={4},
  pages={7441},
  year={2014},
  publisher={Nature Publishing Group}
}

@article{simultaneouscomplexusingphaseslmnature,
  title={Simultaneous shaping of amplitude and phase of light in the entire output plane with a phase-only hologram},
  author={Wu, Liang and Cheng, Shubo and Tao, Shaohua},
  journal={Scientific reports},
  volume={5},
  pages={15426},
  year={2015},
  publisher={Nature Publishing Group}
}

\end{document}